\documentclass[onecolumn]{IEEEtran}
\usepackage{amsmath,amsfonts}
\usepackage{array}
\usepackage[caption=false,font=normalsize,labelfont=sf,textfont=sf]{subfig}
\usepackage{textcomp}
\usepackage{stfloats}
\usepackage{url}
\usepackage{verbatim}
\usepackage{graphicx}
\usepackage[noadjust]{cite}
\usepackage{graphicx}
\usepackage{amsthm}
\usepackage[table]{xcolor}
\usepackage{algorithm}
\usepackage{algpseudocode}
\usepackage{tikz}
\usepackage{amssymb}
\usepackage{mathtools}
\usepackage{overpic}
\usepackage{algorithm}
\usepackage{algpseudocode}

\DeclarePairedDelimiter\abs{\lvert}{\rvert}

\DeclarePairedDelimiter\ceil{\lceil}{\rceil}
\DeclarePairedDelimiter\floor{\lfloor}{\rfloor}
\DeclarePairedDelimiter\parenv{\lparen}{\rparen}

\DeclarePairedDelimiter\set{\{}{\}}

\theoremstyle{plain}
\newtheorem{theorem}{Theorem}
\newtheorem{lemma}[theorem]{Lemma}

\newtheorem{corollary}[theorem]{Corollary}
\newtheorem{definition}[theorem]{Definition}
\newtheorem{example}[theorem]{Example}
\newtheorem{remark}[theorem]{Remark}

\renewcommand{\leq}{\leqslant}

\renewcommand{\geq}{\geqslant}

\newcommand{\cC}{\mathcal{C}}

\newcommand{\cF}{\mathcal{F}}

\newcommand{\cO}{\mathcal{O}}

\newcommand{\cS}{\mathcal{S}}

\newcommand{\F}{\mathbb{F}}
\newcommand{\C}{\mathbb{C}}

\newcommand{\eqdef}{\triangleq}

\DeclareMathOperator{\supp}{supp}
\DeclareMathOperator{\Tr}{Tr}
\DeclareMathOperator{\LC}{LC}
\DeclareMathOperator{\lcm}{lcm}
\DeclareMathOperator{\ord}{ord}
\DeclareMathOperator{\BCH}{BCH}
\DeclareMathOperator{\Melas}{Melas}

\begin{document}

\title{On the Burst-Covering Radius of Binary Cyclic Codes}

\author{
Gabriel~Sac~Himelfarb,~\IEEEmembership{Graduate Student Member,~IEEE} and Moshe~Schwartz,~\IEEEmembership{Fellow,~IEEE}%
\thanks{This paper was submitted in part to the IEEE International Symposium on Information Theory 2026.}%
\thanks{Gabriel Sac Himelfarb is with the Department of Electrical and Computer Engineering, McMaster University, Hamilton, ON, L8S 4K1, Canada (e-mail: sachimeg@mcmaster.ca).
}%
\thanks{Moshe Schwartz is with the Department of Electrical and Computer Engineering, McMaster University, Hamilton, ON, L8S 4K1, Canada (e-mail: schwartz.moshe@mcmaster.ca).}
}

\maketitle

\begin{abstract}
We define and study burst-covering codes. We provide some general bounds connecting the  parameters of a code with its burst-covering radius. We then provide stronger bounds on the burst-covering radius of cyclic codes, by employing linear-feedback shift-register (LFSR) sequences. For the case of BCH codes we prove a bound on pattern frequencies in LFSR sequences, and using this tool, we  bound the burst-covering radius of binary primitive BCH codes and Melas codes. We then present an efficient burst-covering algorithm for cyclic codes. Finally, we present a bound on the critical exponent of cyclic codes based on the burst-covering radius.
\end{abstract}


\section{Introduction}

\IEEEPARstart{E}{rror-correcting} codes and their geometric counterparts, covering codes, have a long and rich history of research (e.g.,~\cite{MacSlo78,CohHonLitLob97}). From a geometric perspective, while the former pack the space with error balls, the latter cover it. In their intersection lie perfect codes, that manage to tile the space with error balls.

Many families of codes have been studied both for their error-correction capabilities, as well as their covering parameters. A partial list of those contains MDS codes, cyclic codes (including BCH, dual BCH, Melas and Zetterberg codes), Reed-Muller codes, and of course, perfect codes (for details, see for example~\cite{MacSlo78,CohHonLitLob97}).

However, one family of codes is conspicuously missing from this list -- codes for burst errors. A $b$-burst error is an error pattern all of whose erroneous symbols are confined to a contiguous block of $b$ positions. From an error-correction perspective, such codes are motivated by the existence of bursty channels, that tend to group together erroneous positions. Burst-correcting codes have a history of research almost as long as that of error-correcting codes, starting with~\cite{Abr59,Fir59,ElsSho62}. The burst-correcting capability of cyclic codes was studied in \cite{Matt1980DeterminingTB}, where it was characterized in terms of the shortest feedback shift register generating certain sequences. Some cyclic burst-correcting codes are optimal, in the sense that they have the closest possible integer parameters to those dictated by the ball-packing bound, making them almost perfect. Among these we mention the cyclic codes of~\cite{AbdMcEOdlTil86,Abd88}, and the cyclic two-dimensional codes of~\cite[Construction A]{SchEtz05}. To the best of our knowledge, only one construction of perfect burst-correcting codes is known~\cite{Etz01b}, for binary codes with burst length $b=2$ \footnote{There is a subtlety in the definitions of optimal and perfect burst-correcting codes: while optimal codes consider bursts cyclically, perfect codes do not.}. 

In this work, we introduce and study \emph{burst-covering codes} for the first time, the natural geometric counterpart to burst-error-correcting codes in the context of covering problems. We are not aware of any previous works in the literature addressing this gap. 

Burst-covering codes do not only fill a void in the theory of binary codes, but may also find use in data storage scenarios. Many applications require the implementation of database queries whose return value is a linear combination of database items with coefficients supplied by the user. Private information retrieval (PIR) protocols~\cite{ChoGolKusSud98}, partial-sum queries~\cite{ChazeRosen89} and more recently, certain machine-learning inference implementations~\cite{RamRavTam24} that are based on ideas from generalized covering radii~\cite{EliFirSch21a,EliWeiSch22}, among others, all employ such queries. Traditionally, covering codes are used to answer such linear queries. This relies on the fact that any column vector can be obtained as a linear combination of at most $R$ columns of a parity-check matrix for the code, where $R$ is the code's covering radius. This guarantees a bounded access complexity, speeding up the computation of the answer to the linear query.

However, in some media types, a significant component in the time to answer a query is not only the number of items that need to be accessed, but also their spatial location (e.g., the seek time in HDDs). Such systems perform best when accessing contiguous blocks of data, as opposed to scattered random-access patterns. This resulted in a surge of interest in codes that take into account access patterns, and prioritize access in contiguous blocks~\cite{wu2021achievable,chee2024repairing,YuYuaSch25}. In the context of our work, as we will prove, the parity-check matrix $H$ of a linear $b$-burst-covering code  satisfies the property that any column vector can be obtained as a linear combination of a window of at most $b$ consecutive columns of $H$. Thus, employing a burst-covering code can address locality issues in computing linear database queries.

Our main contributions are as follows: We first define burst-covering codes and derive basic bounds on their parameters. We then focus on the study of the burst-covering radius of binary cyclic codes. This is motivated by the fact that cyclic codes have a rich structure, they contain some very useful code families (e.g., BCH codes), and the fact that we have almost perfect cyclic burst-correcting codes~\cite{AbdMcEOdlTil86,Abd88}. As we later show, the study of cyclic burst-covering codes is closely related to the analysis of pattern frequencies in linear-feedback shift-register (LFSR) sequences. We employ classic known bounds for the number of occurrences of subwords in these sequences, but for the relevant case of BCH codes we show that these fail to provide significant results. We thus prove a result on pattern frequencies which is meaningful for BCH codes, and is of independent interest for the study of LFSR sequences. We use this to bound the burst-covering radius of binary BCH codes, and the closely related Melas codes. We also show that our analysis naturally gives rise to an efficient covering algorithm for binary cyclic codes. Finally, we present a new bound on the critical exponent of binary cyclic codes based on the burst-covering radius of the dual code.

The paper is organized as follows. Section~\ref{sec:prelim} gives the necessary notation and known results to be used later. In Section~\ref{sec:becc} we define general linear burst-covering codes, and provide some bounds on their parameters. We then study cyclic burst-covering codes in Section~\ref{sec:cyclic}, and further focus on the burst-covering radius of BCH codes in Section~\ref{sec:bch}. An efficient covering algorithm is described in Section~\ref{sec:algo}. The connection to the critical exponent of linear codes is presented in Section \ref{sec:criticalexp}. We conclude in Section~\ref{sec:conc} with a summary of the results and some open questions.

\section{Preliminaries}
\label{sec:prelim}

Let $\F_q$ denote the finite field of size $q$, and $\F_q^*\eqdef \F_q\setminus\set{0}$. We use $\F_q^n$ to denote the set of vectors of length $n$ with entries from $\F_q$, and similarly, $\F_q^{r\times n}$ to denote the set of $r\times n$ matrices with entries from $\F_q$. Vectors will be usually denoted with a lower-case letter, whereas matrices with upper-case ones. Whether a vector is a row or column vector will be understood from the context. We shall usually index entries from $0$, i.e., a vector $v\in\F_q^n$ will be denoted by $v=(v_0,v_1,\dots,v_{n-1})$. The support of a vector $v$ is defined as
\[
\supp(v) \eqdef \set*{ 0\leq i\leq n-1 : v_i\neq 0}.
\]

An $[n,n-r]_q$ linear code, $\cC$, is an $(n-r)$-dimensional space, $\cC\subseteq\F_q^n$. We say $r$ is the redundancy of the code\footnote{It is also common to denote $k=n-r$ to be the dimension of the code, though for convenience, we shall mainly use the code's redundancy $r$.}. By convention, we can specify the code $\cC$ through a parity-check matrix $H\in\F_q^{r\times n}$, such that $c\in \cC$ if and only if $Hc=0$, i.e., $\cC=\ker(H)$. Note that a code may have more than one parity-check matrix. Vectors of the form $Hv$, $v\in\F_q^n$, are called syndromes. Since $H$ is full rank, the set of all syndromes is $\F_q^r$. The dual code of $\cC$, denoted $\cC^\perp$, is the linear code spanned by the rows of $H$.

Given a vector $v=(v_0,\dots,v_{n-1})\in\F_q^n$, a cyclic shift of $v$ is the vector $(v_{n-1},v_0,v_1,\dots,v_{n-2})$. A linear code $\cC$ is said to be cyclic if $c\in \cC$ implies the cyclic shift of $c$ is also in $\cC$. Denote $\F_q[X]$ the set of polynomials in the unknown $X$, with coefficients from $\F_q$. With any vector $v$ we associate the polynomial $v(X)=\sum_{i=0}^{n-1} v_i X^i$. It is well known~\cite{MacSlo78}, that an $[n,n-r]_q$ cyclic code $\cC$ is an ideal in the ring of polynomials $\F_q[X]/(X^n-1)$. There exists a unique generator polynomial, $g(X)\in\F_q[X]$, $\deg(g(X))=r$, such that $c(X)\in\cC$ if and only if $c(X)=u(X)g(X)$, for some $u(X)\in\F_q[X]$, $\deg(u(X))\leq n-r-1$. The roots of $g(X)$ (in its splitting field) are called the roots of the code $\cC$. The commonly studied case is that of $g(X)$ having only simple roots (i.e., no repeated root). This is guaranteed, for example, when $\gcd(n,q)=1$.

\subsection{Linear-feedback shift registers (LFSRs)}

In this section we recall basic definitions and results on linear-feedback shift-register sequences and Galois-mode linear-feedback shift registers. For simplicity of presentation, the treatment in this and subsequent sections will be limited to binary sequences. In the case of larger fields, sign considerations need to be taken into account. 

\begin{definition}
Given a polynomial $f\in\F_q[X]$, its \emph{order} (also known as \emph{exponent}) is the least positive integer $n$ such that $f|X^n-1$. 
\end{definition}

\begin{definition}
    Given a field extension $\F_{q^t}$ of $\F_q$, the \emph{trace} map, $\Tr_{\F_{q^t}/\F_q}:\F_{q^t}\rightarrow \F_q$, is defined as
    \[
    \Tr_{\F_{q^t}/\F_q}(\alpha)=\alpha+\alpha^q+\dots+\alpha^{q^{t-1}}.
    \]
    If the field extension is clear from the context, we will drop the subscript $\F_{q^t}/\F_q$.
\end{definition}

\begin{definition}
    Given a polynomial of degree $r$, $g(X)=\sum_{i=0}^{r-1}m_i X^i +X^r\in \F_2[X]$, we define an LFSR sequence of order $r$ and connection polynomial $g$ (sometimes called, characteristic polynomial) as a sequence $(a_k)_{k\geq 0}$ which satisfies the linear recurrence
\[
a_k = \sum_{i=0}^{r-1}m_i a_{k-r+i}
\]
for all $k\geq r$. The elements $a_0,\dots,a_{r-1}$ are called the initial conditions of the sequence. We say $g$ is the \emph{minimal} connection polynomial in case $(a_k)_{k\geq 0}$ does not satisfy any linear recursion of smaller order. 
\end{definition}

The following theorem summarizes results from \cite{GorKlap12} and \cite{Gol67}:

\begin{theorem}\label{thm:LFSRtracecharacterization}
    \begin{itemize}
        \item [1)] If $g\in\F_2[X]$ is irreducible over $\F_2$, and $\alpha\in \F_{2^r}$ is a root of $g$, there exists some $\beta\in \F_{2^r}$, determined by the initial conditions $a_0,\dots,a_{r-1}$, such that for all $k\geq 0$,
        \[
        a_k = \Tr(\beta \alpha^k),
        \]
        where $\Tr:\F_{2^r}\rightarrow \F_2$ is the trace map from $\F_{2^r}$ to $\F_2$. If the sequence is non-zero, then its minimal period is  $\ord(g)$.
      
        \item [2)] If $g$ factors into distinct irreducible polynomials, $g=\prod_{i=1}^e g_i$, of degrees $d_1,\dots, d_e$ respectively, then for all $k\geq 0$,
        \[
        a_k =\sum_{i=1}^e \Tr_{\F_{2^{d_i}}/\F_2}(\gamma_i \alpha_i^k) ,
        \]
        where $\alpha_i\in \F_{2^{d_i}}$ is a root of $g_i$ and $\gamma_i\in \F_{2^{d_i}}$ $1\leq i\leq e$.

        If $g$ is the minimal polynomial of the sequence, then the minimal period is equal to $\lcm\set{\ord(g_i): 1\leq i\leq e}$.
        
        In the particular case when $d_1=\dots=d_e=d$, we have
        \[
        a_k = \Tr_{\F_{2^{d}}/\F_2}\parenv*{\sum_{i=1}^e \gamma_i \alpha_i^k}.
        \]
    \end{itemize}
\end{theorem}

\subsection{Galois-mode LFSRs}
We now introduce Galois-mode linear-feedback shift registers. We refer the reader to~\cite{GorKlap12}, although our treatment differs slightly. 

Given a polynomial of degree $r$, $g(X)=\sum_{i=0}^{r-1}m_i X^i +X^r\in \F_2[X]$, we define the Galois-mode LFSR of length $r$ and connection polynomial $g$ as a sequence generator with states of the form $(f_0,f_1,\dots,f_{r-1})\in \F_2^r$, and whose state transition is given by:
\[
(f_0,f_1, \dots, f_{r-1}) \longrightarrow (m_{0}f_{r-1},f_0+m_1 f_{r-1}, f_1+m_2 f_{r-1},\dots, f_{r-2}+m_{r-1} f_{r-1}).
\]
The output of the LFSR is the sequence of elements $f_{r-1}$ from each state.

\begin{theorem}\label{thm:galois}
    Given a Galois-mode LFSR of length $r$ and connection polynomial $g$ as above,
    \begin{itemize}
        \item[1)] If we identify a state $(f_0,\dots,f_{r-1})$ with the polynomial $f(X)=\sum_{i=0}^{r-1} f_i X^i$, then the sequence of states corresponds to the sequence of polynomials $(X^kf \pmod g)_{k\geq 0}$, where $f$ is the initial state.

        \item [2)] The output $(a_k)_{k\geq 0}$ satisfies the linear recurrence
        \[
        a_k = m_{r-1}a_{k-1}+m_{r-2}a_{k-2}+\dots+m_0a_{k-r},
        \]
        for every $k\geq r$, which means that it can also be obtained as the output of an LFSR with connection polynomial $g$. 
    \end{itemize}
\end{theorem}

\begin{IEEEproof}
    \begin{itemize}
        \item [1)] Consider a state $f$. If $\deg(f)<r-1$, or equivalently, $f_{r-1}=0$, then $Xf\pmod g = Xf$, and the coefficients obey the state transition $(f_0,\dots,f_{r-1})\longrightarrow (0,f_0,\dots,f_{r-2})$. 

        If $\deg(f)=r-1$, then $Xf \pmod g = Xf+g$, and the new coefficients are $(m_0,f_0+m_1,\dots,f_{r-2}+m_{r-1})$.

        In both cases we see that the change of polynomial coefficients obeys the Galois-mode LFSR state transition.

        \item [2)] From the proof of 1) we can see that $X^{k+1}f \pmod g = X\cdot(X^kf\pmod g)+ a_k g$. It follows inductively that 
        \[
        X^kf \pmod g= X^kf+g(a_0X^{k-1}+a_1X^{k-2}+\dots+a_{k-1})
        \]
        By looking at the coefficient of $X^{r-1}$ we get the desired recurrence.
    \end{itemize}
\end{IEEEproof}

\begin{theorem}\label{thm:galoisconsecutivezeros}
    Given $g\in \F_2[X]$ of degree $r$, consider the Galois-mode LFSR with connection polynomial $g$ as above and initial load $f\neq 0$. Denote by $(a_k)_{k\geq 0}$ the output sequence. Then
    \[
    \deg(X^kf \pmod{g}) = r-1-\max \set*{j: a_k=0, a_{k+1}=0, \dots, a_{k+j-1}=0}, 
    \]
    where the maximum is taken to be $0$ if $a_k=1$.
\end{theorem}
\begin{IEEEproof}
    By the definition of the output sequence of the Galois-mode LFSR and by Theorem \ref{thm:galois}, $a_k$ is the coefficient of $X^{r-1}$ in $X^k f \pmod g$. Thus, if $a_k=1$, $\deg(X^kf \pmod g)=r-1$.

    If $a_k=a_{k+1}=\dots = a_{k+j-1}=0$ and $a_{k+j}=1$, then $\deg(X^kf \pmod g)$, $\deg(X^{k+1}f \pmod g), \dots, \deg(X^{k+j-1}f \pmod g)$ are all less than $r-1$, and $\deg(X^{k+j}f \pmod g)=r-1$. Thus we have
    \[X^{k+1}f\pmod g = X\cdot (X^kf \pmod g), \quad \dots \quad , X^{k+j}f\pmod g = X^{j}\cdot (X^kf \pmod g).\]
    From this last equality, we deduce $r-1=j+ \deg(X^kf \pmod g)$.
\end{IEEEproof}

\begin{figure}
\begin{center}
\begin{overpic}[scale=0.4]
{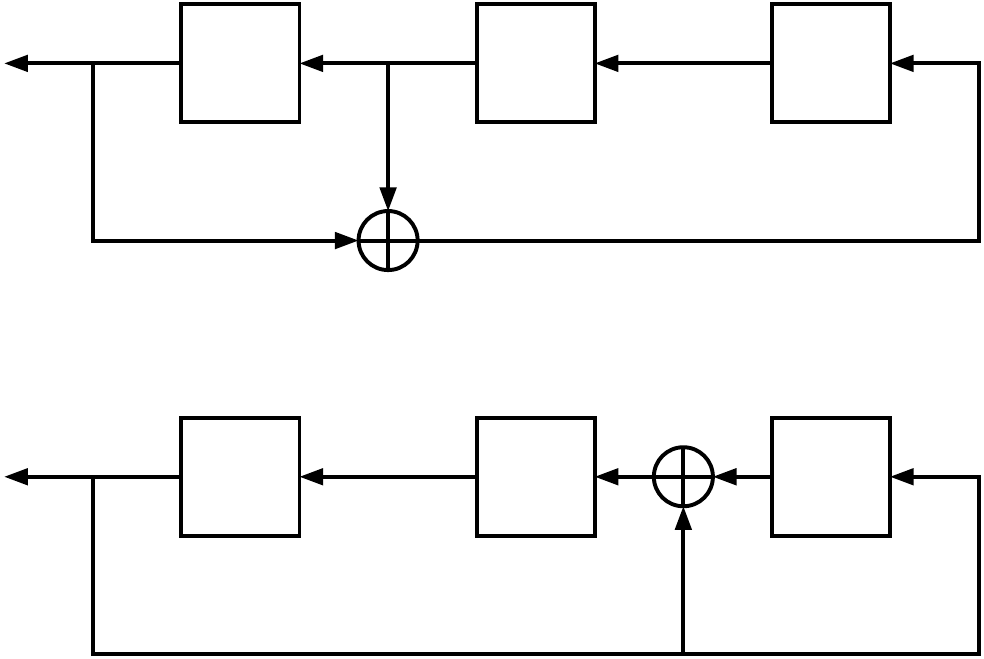}
\put(0,50){(a)}
\put(0,10){(b)}
\end{overpic}
\end{center}
\caption{Generating a binary sequence with connection polynomial $g(X)=1+X+X^3$ using (a) an LFSR, and (b) a Galois-form LFSR. The boxes represent flip-flops storing a single bit, and $\oplus$ is an XOR gate. }
\label{fig:lfsr}
\end{figure}

An example of generating a binary LFSR sequence with connection polynomial $g(X)=1+X+X^3$ is shown in Figure~\ref{fig:lfsr}. The figure shows two circuits generating the same sequence: the first a standard LFSR, and the second, a Galois-form LFSR. 

\subsection{Character sums}

\begin{definition}[{{\cite[Chapter 5]{LidNie97}}}]
    An additive character over a finite field $\F_q$ is a function $\chi:\F_q\rightarrow \C$ that satisfies $\chi(x+y)=\chi(x)\cdot \chi(y)$. In other words, it is an homomorphism from the additive group of the field to the multiplicative group of modulus-$1$ complex numbers.
\end{definition}

The canonical additive character of $\F_q$ is defined as
\[
\chi(x)=e^{2\pi i \Tr(x)/p},
\]
where $p$ is the characteristic of the field $\F_q$, and any other additive character can be defined as $\chi_y(x)=\chi(y\cdot x)$. If we choose $y=0$ we obtain the trivial additive character. Most of the results in the following sections are for binary codes. In this case, we shall use the canonical additive character $\chi(x)=(-1)^{\Tr(x)}$.

\begin{theorem}[Weil-Carlitz-Uchiyama bound, \cite{CarlitzUchiyama57}]
\label{thm:weil}
Let $f\in \F_q[X]$ be of degree $n\geq 1$ with $\gcd(n,q)=1$, and let $\chi$ be a non-trivial additive character of $\F_q$. Then
\[
\abs*{\sum_{x\in \F_q}\chi(f(x))}\leq (n-1) q^{1/2}.
\]
\end{theorem}

Theorem~\ref{thm:weil} has been extended to rational functions, i.e., ratios of polynomials. The set of all rational functions in the unknown $X$, and coefficients from $\F_q$, is denoted by $\F_q(X)$.

\begin{theorem}[\cite{MorMor91}]\label{thm:generalizedweilbound}
Let $f(X)=\frac{P(X)}{Q(X)}\in \F_q(X)$ be a rational function over a finite field of characteristic $p$, and $\chi$ a non-trivial additive character of $\F_q$. Suppose $f$ is non-degenerate, in the sense that there does not exist $h\in\overline{\F_q}$ such that $f=h^p-h$, where $\overline{\F_q}$ is the algebraic closure of $\F_q$. Let $s$ be the number of distinct roots of $q$ in $\overline{\F_q}$. Then,
    \[
   \abs*{\sum_{x\in \F_q\setminus\cS}\chi(f(x))}\leq (\max(\deg P, \deg Q)+s'-2)q^{1/2}+\delta,
    \]
     where $\cS$ is the set of poles of $f$;  $s'=s$, $\delta=1$ if $\deg P\leq \deg Q$, and $s'=s+1$ and $\delta=0$ otherwise.
\end{theorem}

\begin{corollary}\label{coro:generalweilconsequence}
    Let $f(X)=\sum_{i=1}^e a_iX^{t_i}+\sum_{i=1}^d b_iX^{-u_i}\in \F_{q}(X)$, where $q=2^m$,  $e,d>0$, and $t_e>\dots>t_1>0$, $u_d>\dots>u_1>0$, are all odd integers, as well as the coefficients $a_i$ and $b_i$ are non-zero. Then, $f$ is non-degenerate and
   \[
    \abs*{\sum_{x\in \F_q^*}\chi(f(x))}\leq (t_e+u_d)q^{1/2}.
    \]
\end{corollary}

\begin{IEEEproof}
    Suppose $f$ were degenerate, and that there existed $h$, a rational function such that $f=h^2-h$. Suppose $h=h_1/h_2$ with $h_1$ and $h_2$ polynomials and $\gcd(h_1,h_2)=1$. We can rearrange the equality as:
    \[
    h_2^2\parenv*{\sum_{i=1}^e a_i X^{t_i+u_d}+\sum_{i=1}^d b_iX^{u_d-u_i}} = X^{u_d}h_1(h_1-h_2).
    \]

    If $h_2=X^i \widetilde{h_2}$, $X\nmid \widetilde{h_2}$, then $\widetilde{h_2}|\text{RHS}$, which implies that $\widetilde{h_2}|h_1-h_2$, which is a contradiction  since $\widetilde{h_2}$ and $h_1$ are coprime, unless $\widetilde{h_2}=1$. Thus, we deduce that $h_2=X^i$ for some $i\geq 0$, and:
    \[
    X^{2i}\parenv*{\sum_{i=1}^e a_iX^{t_i+u_d}+\sum_{i=1}^d b_iX^{u_d-u_i}} = X^{u_d}h_1(h_1-X^i).
    \]
    Since $X$ does not divide the second factor in the LHS, we have that $2i\geq u_d$. However, since $u_d$ is odd, we deduce $2i>u_d$, which implies that $X|h_1(h_1-X^i)$. If $i\geq 1$, we deduce $X|h_1$, which is a contradiction since $\gcd(h_1,h_2)=1$. If $i=0$, then $X\nmid \text{LHS}$ and $X|\text{RHS}$ (since $u_d\geq 1$), which is a contradiction. This means $f$ is non-degenerate.
    
    To conclude, $f=(\sum_{i=1}^e a_i X^{t_i+u_d}+\sum_{i=1}^d b_i X^{u_d-u_i})/X^{u_d}$, so the number of distinct roots of the denominator is $1$ and the degree of the numerator is strictly greater than the degree of the denominator. Using the same notations as in Theorem \ref{thm:generalizedweilbound} we have $s'=2$, $\delta=0$, and the claim follows.

\end{IEEEproof}

\section{Burst-Covering Codes}
\label{sec:becc}

Linear covering codes have several equivalent definitions. In particular, a geometric definition shows how the space is covered by error balls surrounding the codewords, and an algebraic definition shows how syndromes are covered by linear combinations of columns from a parity-check matrix for the code. We use these two approaches to define burst-covering codes.

Our first definition is a geometric one. It calls for the definition of a burst-error-ball:

\begin{definition}
    Given $x\in \F_q^n$, $b>0$ an integer, and an index $0\leq i<n$, define:
    \[
    B_b(x,i) \eqdef \set*{y\in \F_q^n : \supp(y-x)\subseteq \set*{i,i+1,\dots, i+b-1}},
    \]
    where indices are considered modulo $n$. In other words, $B_b(x,i)$ is the set of vectors that differ from $x$ inside a window of $b$ consecutive positions starting at $i$, where indices are viewed cyclically.  
    
    We define the \emph{$b$-burst ball} of radius $b$, centered at $x\in \F_q^n$, as:
    \[
    B_b(x) \eqdef \bigcup_{i=0}^{n-1} B_b(x,i)
    \]
\end{definition}

In the Hamming metric, the Hamming distance between two vectors $x$ and $y$ is the smallest radius of a ball centered at $x$, that contains $y$. It is thus tempting to use the same approach with burst balls. However, we note that
\[
d(x,y)=\min \set*{b: y\in B_b(x)}
\]
is not a metric, as it does not satisfy the triangle inequality. For example, 
\[3=d(10100,00000) > d(10100,10000)+d(10000,00000)=1+1=2.\]

Even so, we use the burst balls to provide a geometric definition of burst-covering codes:

\begin{definition}\label{def:burstcode}
    We say $\cC\subseteq \F_q^n$ is a $b$-burst covering code if 
    \[
    \bigcup_{x\in \cC} B_b(x) = \F_q^n.
    \]
\end{definition}

Similarly to what occurs in the case of regular covering codes, we can give an equivalent definition for linear codes based on their parity-check matrix:

\begin{theorem}\label{thm:equivdefinition}
Let $\cC$ be a linear $[n,n-r]_q$ code, and let $H\in\F_q^{r\times n}$ be a parity-check matrix for the code. Then $\cC$ is a $b$-burst covering code if and only if every column vector $z\in\F_q^{r}$ can be obtained as a linear combination of $b$ (cyclically) consecutive columns of $H$. Additionally, $b$ does not depend on the choice of $H$.
\end{theorem}

\begin{IEEEproof}
Given any $z\in \F_q^{r}$, take any $y\in \F_q^{n}$ such that $Hy=z$. If $\cC$ is $b$-burst covering, there exists $c\in \cC$ such that $\supp(y-c)\subseteq \set{i,i+1,\dots,i+b-1}$ for some index $i$. Then $H(y-c)=Hy=z$ is a linear combination of $b$ consecutive columns starting in position $i$.

For the converse, given any $y\in \F_q^n$, $Hy\in \F_q^{r}$ has to be a linear combination of at most $b$ consecutive columns of $H$. Thus, there exists a vector $w\in\F_q^n$ with $\supp(w)\subseteq \set{i,i+1,\dots,i+b-1}$ for some $i$, such that $Hy=Hw$. Thus, $H(y-w)=0$, and $y-w\in \cC$. Then $y\in B_b(y-w)$, and the code is $b$-burst covering.
\end{IEEEproof}

The remainder of this work will focus on linear burst-covering codes, and thus we will only use the equivalent formulation from Theorem~\ref{thm:equivdefinition} as our definition. We will also consider burst-covering codes where the columns in the parity-check matrix are not viewed cyclically. In that case we will make the distinction explicit. 

\begin{definition}
The \emph{burst-covering radius} of a code $\cC$ is the least integer $b$ such that $\cC$ is a $b$-burst covering code. The burst-covering radius of a full-rank matrix $H\in\F_q^{r\times n}$, $r\leq n$, is the least integer $b$ such that any column vector $z\in \F_q^{r}$ can be obtained as a linear combination of $b$ consecutive columns of $H$.  
\end{definition}

\begin{remark}
    Unlike other code parameters such as the minimum distance or the covering radius, the burst-covering radius is not necessarily invariant under permutations of the bit positions. However, the burst-covering radius of a matrix $H$ is invariant under row operations, since these do not alter the code defined by $H$. 
\end{remark}

\begin{example}
Consider the $[8,4,4]_2$ extended binary Hamming code, with parity-check matrix
\[
H=\begin{pmatrix}
1 & 1 & 1 & 1 & 1 & 1 & 1 & 1\\
0 & 0 & 0 & 0 & 1 & 1 & 1 & 1\\
0 & 0 & 1 & 1 & 0 & 0 & 1 & 1\\
0 & 1 & 0 & 1 & 0 & 1 & 0 & 1\\
\end{pmatrix}.
\]
It is well known, and easy to see, that the regular covering radius of the code is $2$, since any column vector from $\F_2^4$ can be shown to be the sum of at most two columns from $H$, and $2$ is the smallest number with this property.

However, the burst-covering radius of the code is $4$, since for example, the syndrome $s=(0,1,0,1)$ may be obtained as the sum of the second and fifth columns of $H$, i.e., a burst of length $4$, and no shorter burst produces $s$. 

By permuting the columns of $H$ we can define 
\[
H' = \begin{pmatrix}
1 & 1 & 1 & 1 & 1 & 1 & 1 & 1\\
0 & 0 & 1 & 1 & 1 & 0 & 0 & 1\\
0 & 0 & 0 & 1 & 1 & 1 & 1 & 0\\
0 & 1 & 0 & 0 & 1 & 1 & 0 & 1\\
\end{pmatrix},
\]
whose burst-covering radius is $3$. This shows how the burst-covering radius is not invariant under coordinate permutations.
\end{example}

\begin{theorem}
    Let $\cC$ be an $[n,n-r]_q$ $b$-burst-covering code. Then, 
    \begin{itemize}
        \item [a)] If bursts are considered cyclically and $b\geq 2$:
        \begin{equation}\label{eq:bound1}
        n\geq \frac{q^{r-b+1}-1}{q-1}+1   
        \end{equation}

        \item [b)] If bursts are considered non-cyclically: 
        \begin{equation}\label{eq:bound2}
            n\geq \frac{q^{r-b+1}-1}{q-1}+b-1
        \end{equation}
        with equality if and only if all non-zero vectors occur exactly once as a $b$-burst linear combination of columns of $H$. 
    \end{itemize}
\end{theorem}

\begin{IEEEproof}
    a) There are $n\cdot (q-1)\cdot  q^{b-1}$ linear combinations of $b$ consecutive columns: there are $n$ options for the rightmost column with non-zero coefficient in the combination, $q-1$ options for its coefficient, and $q^{b-1}$ choices of coefficients for the previous $b-1$ columns. Since the code is $b$-burst covering, we have that
    \[
    n(q-1)q^{b-1}\geq q^{r}-1,
    \]
    which implies
    \[
    n\geq \frac{1}{q^{b-1}}\sum_{i=0}^{r-1}q^i = \sum_{i=0}^{b-2}\frac{1}{q^{b-1-i}}+\sum_{i=0}^{r-b}q^i = \frac{q^{r-b+1}-1}{q-1}+\sum_{i=0}^{b-2}\frac{1}{q^{b-1-i}}.
    \]
    Since $n$ is an integer, the claimed bound follows.

    b) For each $0\leq i\leq b-2$, there are $(q-1)q^i$ linear combinations where $i$ is the rightmost column with a non-zero coefficient. There are $(n-b+1)(q-1)q^{b-1}$ other linear combinations. Thus:
    \[
    (n-b+1)(q-1)q^{b-1}+\sum_{i=0}^{b-2} (q-1)q^i \geq q^{r}-1.
    \]
    After rearranging,
    \[
    (n-b+1)(q-1)\geq q^{r-b+1}-1 ,
    \]
    and the desired bound follows. 
\end{IEEEproof}

In case \eqref{eq:bound2} is attained with equality, the resulting code is in fact \emph{perfect}, i.e., simultaneously $b$-burst correcting and $b$-burst covering. To the best of our knowledge, only a single construction is known for such a code, with $q=2$ and $b=2$ (see \cite{Etz01b}). 

For binary codes and burst sizes greater than $2$, we can improve bound \eqref{eq:bound1}:
\begin{theorem}\label{th:improv}
If $\cC$ is an $[n,n-r]_2$ $b$-burst-covering code with $b\geq 3$ and $r \geq 2$, then in the cyclical-burst case we have
    \begin{equation}\label{eq:bound3}
        n\geq 2^{r-b+1}+1.
    \end{equation}
\end{theorem}
\begin{IEEEproof}
    Let $H\in\F_2^{r\times n}$ be a parity-check matrix for the code. Suppose that $n= 2^{r-b+1}$. Then, $n2^{b-1}=2^{r}$, which implies that one of the following occurs:
    \begin{itemize}
        \item [a)] all syndromes, including zero, are obtained exactly once as a linear combination of $b$ consecutive columns of $H$; or
        \item [b)] all non-zero syndromes are linear combinations of at most $b$ consecutive columns in exactly one way, except that exactly one is repeated.
    \end{itemize}
    
    To see that a) cannot occur, notice that no column of $H$ can be zero (otherwise, adding this column to an adjacent one would repeat a syndrome), and if zero is the sum of $2$ or more columns in a window of size $b$, this sum can be split into two equal sums, which would mean there is a repeated syndrome.

    To see that b) is also impossible, let us prove that if $b>2$ then the number of linear combinations in which each column participates is even. In the matrix $H$, consider a column $x$, the column to its right $y$, and to its left $z$. We can define a bijection in the set of linear combinations containing $x$ in the following way: 
    \begin{itemize}
        \item If $y$ is in the linear combination, remove it.
        \item If $y$ is not in the linear combination and can be added preserving the window size less than or equal to $b$, then add it. 
        \item If $y$ is not in the linear combination and cannot be added:
            \subitem - If $z$ is not in the linear combination, then add it (this is always possible because $y$ cannot be added, so the window is guaranteed to go to the left).
            \subitem - If $z$ is in the linear combination, remove it.
    \end{itemize}

    The only possible way this operation does not yield a bijection is if removing $z$ (from a combination to which $y$ cannot be added) yielded a linear combination to which $y$ can be added. This can only happen if the combination was $x+z$. But if $y$ could not be added to $x+z$, this means that $b=2$.

     Let $h_1,h_2,\dots,h_n$ be the columns of $H$, and consider all valid distinct linear combinations of $b$ consecutive columns $\sum_{j=0}^{b-1}c_j h_{i+j}$. When adding up all these combinations, each column vector appears an even number of times as a summand, and the result is $0$. On the other hand, if $w$ is the repeated linear combination value, $w+\sum_{v\in \F_2^r}v=w$, which implies $w=0$, a contradiction.
\end{IEEEproof}

\section{Burst-Covering Radius of Binary Cyclic Codes}
\label{sec:cyclic}

We will restrict our study to binary cyclic codes with no repeated roots. Consider a cyclic code with generator polynomial $g(X)=\prod_{i=1}^e g_i(X)$, where $g_i\in \F_2[X]$, $1\leq i\leq e$, are distinct irreducible polynomials of degrees $d_1,d_2, \dots, d_e$ respectively, and $r=\deg(g)=\sum_{i=1}^e d_i$ is the redundancy of the code. Let $\alpha_i\in\F_{2^{d_i}}$ be a  root of $g_i$ for each $i$. Then, a possible parity-check matrix for the code is
\begin{equation}\label{eq:cyclicH}
H = \begin{pmatrix}
    1 & \alpha_1 & \alpha_1^2 & \dots & \alpha_1^{n-1}\\
    1 & \alpha_2 & \alpha_2^2 & \dots  & \alpha_2^{n-1}\\
    \vdots & \vdots & \vdots & \ddots & \vdots \\
    1 & \alpha_e & \alpha_e^{2} & \dots & \alpha_e^{n-1}
\end{pmatrix},    
\end{equation}
where the $i$-th row should be interpreted as $d_i$ rows, after replacing each $\alpha_i^j$ by its binary vector representation over $\F_2^{d_i}$ (after a choice of some basis).

Given a window of $t$ consecutive columns $h_i, h_{i+1},\dots, h_{i+t-1}$ in $H$, we will identify a linear combination $\sum_{j=0}^{t-1}c_jh_{i+j}$ with the pair $(i,f)$, where $f$ is the polynomial $f(X)=\sum_{j=0}^{t-1}c_jX^j\in \F_2[X]$. Observe that 
\begin{equation}\label{eq:lincombvalue}
\LC(i,f)\eqdef \sum_{j=0}^{t-1}c_jh_{i+j} = \begin{pmatrix}
    \alpha_1^i f(\alpha_1) \\
    \alpha_2^{i} f(\alpha_2) \\
    \vdots \\
    \alpha_e^{i} f(\alpha_e)
\end{pmatrix}.    
\end{equation}
Given a polynomial $f\in \F_2[X]$, we denote by $\LC(f)$ the set
\[
\LC(f) \eqdef \set*{\LC(i,f): 0\leq i\leq n-1}.
\]

We begin by giving crude upper and lower bounds:
\begin{theorem}\label{thm:basicboundcyclic}
    The burst-covering radius, $b$, of a cyclic $[n,n-r]_q$ code with generator polynomial $g$ as defined above, satisfies:
    \[
    r-\min_{1\leq i\leq e} d_i+1\leq b\leq r.
    \]
\end{theorem}

\begin{IEEEproof}
    Without loss of generality, suppose $d_e=\min_i d_i$. Thanks to~\eqref{eq:lincombvalue}, we know that any linear combination $(i,f)$ that yields the syndrome $(0, \dots, 0, 1)$, needs to satisfy $f(\alpha_1)=\dots=f(\alpha_{e-1})=0$. Since $g_1, \dots, g_{e-1}$ are pairwise coprime, $\prod_{i=1}^{e-1}g_i|f$, and so $\deg(f)\geq \sum_{i=1}^{e-1}d_i= \deg(g)-d_e$. We also note that $\LC(i,Xf)=\LC(i+1,f)$, so we may assume $i$ is set such that $X\nmid f$. But that the coefficients of the linear combination described by $f$ satisfy $c_0\neq 0$ and $c_{\deg(f)}\neq 0$, which means that at least $\deg(g)-d_e+1$ consecutive columns are needed. 
    
    Again by~\eqref{eq:lincombvalue}, any linear combination of columns $(i,f)$ which is equal to $0$ has to satisfy $\prod_{i=1}^e g_i|f$, thus if $f\neq 0$, $\deg(f)\geq \deg(g)$. So, any non-trivial linear combination of columns which is $0$ requires at least $\deg(g)+1$ consecutive columns. This means that the first $\deg(g)$ columns of $H$ are linearly independent, and a basis for $\F_2^{r}$, which implies that a window of size $\deg(g)$ suffices.
\end{IEEEproof}

\begin{remark}
    If $\deg(g_i)=1$ for some $1\leq i\leq e$, which is equivalent to having a parity-check bit, then the upper and lower bounds in Theorem \ref{thm:basicboundcyclic} coincide, and the burst-covering radius is equal to $\deg(g)$. 
\end{remark}

\begin{remark}
The following bound is the analogous of~\eqref{eq:bound1} in the context of binary burst-correcting codes:
\[
n\leq 2^{r-b+1}-1.
\]
A $b$-burst-correcting code with parameters $[n,n-r]_2$ which attains this bound is called \emph{optimal} (not to be confused with perfect as defined for bound \eqref{eq:bound2}). Optimal codes have been constructed for different values of the parameters, and their existence in general was proven in \cite{AbdMcEOdlTil86}. For example, in \cite{ElsSho62} it is shown that taking a binary cyclic code with generator polynomial
    \[
    g(X)= (1+X+X^2)f(X),
    \]
with $f$ a primitive polynomial of degree $m=s-2$, $2|m$, $m\geq 4$, and with length $n=2^m-1$, yields an optimal binary cyclic burst-error-correcting code with parameter $b=3$.

According to Theorem \ref{thm:basicboundcyclic}, the burst-covering radius of that code is at least $s-1$. This shows that even an optimal burst-correcting code might have a large discrepancy between its error-correcting parameter $b$ (burst-packing radius) and its burst-covering radius.

\end{remark}

We will now characterize the sets $\LC(f)$:

\begin{lemma}\label{lemma:fxkg}
    For $f_1,f_2\in \F_2[X]$, $\LC(f_1)=\LC(f_2)$ if and only if $f_1(X)\equiv X^k f_2(X) \pmod {g(X)}$ for some $k\geq 0$. 
\end{lemma}

\begin{IEEEproof}
    If $\LC(f_1)=\LC(f_2)$ then there exist $i$ and $k$ such that $\LC(i, f_1)=\LC(i+k, f_2)$. This means that the syndrome corresponding to $X^if_1(X)+X^{i+k} f_2(X) \pmod{X^n-1}$ (interpreted as a vector in $\F_2^n$) is $0$, and thus it is a codeword. We deduce that $g|X^if_1(X)+X^{i+k} f_2(X)$, from which the result follows.
    For the converse, if $f_1(X)\equiv X^k f_2(X) \pmod{g}$, then $f_1(X)+X^k f_2(X) \pmod{X^n-1}$ is a codeword, and thus the syndromes of $X^i f_1(X)$ and $X^{i+k}f_2(X)$ are the same.
\end{IEEEproof}

If we consider $\set{1,X,X^2, \dots, X^{n-1}}$ acting on $\cF_{r}\eqdef \set{f\in \F_2[X]:  \deg(f)<r}$ by multiplication modulo $g$, Lemma \ref{lemma:fxkg} shows that $\LC$ is constant over the orbits. Moreover, the $\LC$ of polynomials in different orbits do not intersect.

From Theorem \ref{thm:basicboundcyclic} and Lemma \ref{lemma:fxkg} we arrive at the following characterization of the burst-covering radius of cyclic codes:

\begin{corollary}\label{coro:radiuscharacterization}
    The burst covering radius $b$ of a binary cyclic code satisfies
    \begin{equation}\label{eq:equivalentdefradius}
        b = \max_{f\in \cF_{r},f\neq 0} \min_{k\geq 0} \deg (X^k f \pmod g)+1.
    \end{equation}
\end{corollary}
Notice that this result holds even if we do not require the roots to be simple, since the proof of Lemma \ref{lemma:fxkg} does not rely on this assumption, and the upper bound from Theorem \ref{thm:basicboundcyclic} is valid for any cyclic code (this can be seen from the standard parity-check matrix constructed from the parity-check polynomial, which has $r$ consecutive linearly independent columns, see for example \cite[Chapter 7]{MacSlo78}).

We will now see that the study of the expression in Corollary \ref{coro:radiuscharacterization} is closely related to the analysis of pattern frequencies in LFSR sequences. More precisely, computing the burst-covering radius of cyclic codes is equivalent to studying the length of runs of consecutive zeros in LFSR sequences with connection polynomial equal to the generator polynomial of the code:

\begin{corollary}\label{coro:equivdefconseczeros}
    The burst covering radius of a binary cyclic code $\cC$ is given by:
    \begin{equation}\label{eq:bfinalequivdefinition}
    b = r- \min_{(a_0,\dots,a_{r-1})\neq 0} Z(g,(a_0, \dots,a_{r-1})),
\end{equation}
where $Z(g,(a_0,\dots,a_{r-1}))$ denotes the maximum length of a  run of zeros in the LFSR sequence with connection polynomial $g$ and initial conditions $a_0, \dots, a_{r-1}$. Alternatively,
\begin{equation*}
    b= r- \min_{c\in \cC^{\perp}, c\neq 0}Z(c),
\end{equation*}
where $Z(c)$ is the maximum length of a run of zeros in $c$ (where runs of zeros in codewords are considered cyclically).
\end{corollary}

\begin{IEEEproof}
    In the Section~\ref{sec:prelim} we recalled the connection between the computation of $X^kf\pmod g$ and a Galois-mode LFSR with connection polynomial $g$. By Theorem~\ref{thm:galoisconsecutivezeros} and \eqref{eq:equivalentdefradius}: 
    \begin{equation*}
    \begin{split}
        b & = \max_{a_0,\dots,a_{r-1}} \min_{k\geq 0}  \parenv*{1+r-1-\max\set{j: a_k=0,\dots, a_{k+j-1}=0}} \\
    & = r- \min_{a_0,\dots,a_{r-1}} \max_{k\geq 0}\max\set{j: a_k=0,\dots, a_{k+j-1}=0}
    \end{split}
\end{equation*}
where the minimum is taken over all possible initial conditions $a_0, \dots, a_{r-1}$ for the LFSR sequence. The LFSR sequences with connection polynomial $g$ are precisely the codewords of the dual code of $\cC$, and the second equation follows.
\end{IEEEproof}

\subsection{Pattern frequencies in LFSR sequences}
We begin this section by recalling a previous result on the number of occurrences of patterns in LFSR sequences:

\begin{theorem}[\cite{Niederreiter86}]
\label{thm:niederreiterfreq}
    Consider a binary LFSR sequence with minimal period $\pi$. Let $g\in \F_2[X]$ be the minimal polynomial of the sequence, and $r=\deg(g)$. Suppose that $s$ is a positive integer less than or equal to the degree of any irreducible factor of $g$ in $\F_2[X]$. Then for any pattern $y\in\F_2^s$, its number of occurrences $N$ in one minimal period of the sequence satisfies
    \begin{equation}\label{eq:niederreiterfreq}
        \abs*{ N-\frac{\pi}{2^s}} \leq \parenv*{1-\frac{1}{2^s}} 2^{r/2}.
    \end{equation}    
\end{theorem}

Theorem \ref{thm:niederreiterfreq} will allow us to study the burst-covering radius of general cyclic codes. However, there is a notable case for which this result is not useful: if $g$ factors into irreducible polynomials of equal degrees, then the lower bound on $N$ that can be derived from \eqref{eq:niederreiterfreq} can be negative, as the following example shows.

\begin{example}
If $g=g_1\cdot g_2$, with $g_1,g_2\in\F_2[X]$, both irreducible of degree $m$, and $g_1$ is primitive, then the period of an LFSR sequence with minimal polynomial $g$ is $\pi=2^m-1$. In that case, the bound of Theorem~\ref{thm:niederreiterfreq} becomes 
\[\abs*{2^sN-(2^m-1)}\leq (2^s-1)2^m.\]
The lower bound we can derive on $N$ is $2^sN\geq 2^m-1-(2^s-1)2^m$, which is negative for any $s\geq 1$.
\end{example}

This problem arises in the study of relevant codes such as long BCH codes. Thus, we introduce the following result, whose proof follows the same line as that of Theorem \ref{thm:niederreiterfreq}, but differs in the application of the Weil-Carlitz-Uchiyama bound. A similar result can be derived for the all-zero pattern from \cite[Theorem V.4]{ShiOzb21}, which is framed in terms of $b$-symbol weight rather than counting LFSR pattern frequencies.

\begin{theorem}\label{thm:patternupperbound}
Consider a connection polynomial $g(X)= \prod_{i=1}^e g_i(X)$, where $g_i\in \F_2[X]$, $1\leq i\leq e$, are distinct irreducible polynomials of the same degree $m$. Let $\alpha$ be a primitive element in the splitting field $\F_{2^m}$ of $g$, and let $t_i\geq 1$ be such that $\alpha^{t_i}$ is a root of $g_i$ for each $1\leq i\leq e$. Further assume that all $t_i$ are odd. Under these conditions, the following holds:
     
Consider any pattern $y\in \F_2^s$ of length $s\leq m$. If $\max_{i=1,\dots,e}t_i\geq 3$, then the number of occurrences $N$ of $y$ in a window of length $2^m-1$ 
(counting occurrences cyclically) of any non-zero LFSR sequence with connection polynomial $g$ as above\footnote{Note that $g$ need not be the minimal polynomial of the sequence.}, satisfies
     \begin{equation}\label{eq:thmpattern}
         \abs*{N-\frac{2^m-1}{2^s}} \leq \parenv*{1-\frac{1}{2^s}}\parenv*{(\max_i t_i -1)2^{m/2}+1}.
     \end{equation}
\end{theorem}

\begin{IEEEproof}
    From Theorem \ref{thm:LFSRtracecharacterization} we know that the sequence is $(a_k=\Tr(\sum_{i=1}^e \gamma_i \alpha^{t_ik}))_{k\geq 0}$, where $\gamma_i\in \F_{2^m}$, $1\leq i\leq e$, determine the initial conditions, and $\Tr:\F_{2^m}\rightarrow \F_2$ is the trace map. We observe that, over the reals, $(1+(-1)^{a+a_k})/2$ is $1$ if $a_k=a$ and $0$ otherwise. Recalling that $\chi(x)=(-1)^{\Tr(x)}$ is the canonical additive character of the field $\F_{2^m}$, we deduce that   
        \[
    \frac{1+(-1)^a\chi (\sum_{i=1}^e \gamma_i \alpha^{t_ik})}{2}
    \]
    is an indicator function of the $k$-th element of the sequence being equal to $a$.

    Thus, the following sum counts the number of occurrences of the pattern $y$ in the first $2^m-1$ bits of the sequence:
    \[
    N = \sum_{k=0}^{2^m-2} \prod_{j=0}^{s-1} \frac{1+(-1)^{y_j}\chi(\sum_{i=1}^e \gamma_i \alpha^{t_i(k+j)})}{2}
    \]
    Making the substitution $x=\alpha^k$, and letting $[s-1]=\set{0,1,\dots,s-1}$ we get
    \begin{equation*}
    \begin{split}
         N = \frac{1}{2^s}\sum_{x\in \F_{2^m}^*} \prod_{j=0}^{s-1} \left(1+(-1)^{y_j}\chi\left(\sum_{i=1}^e \gamma_i \alpha^{t_i j}x^{t_i}\right)\right)   & = \frac{1}{2^s}\sum_{x\in \F_{2^m}^*} \left(1+ \sum_{\substack{J\subseteq [s-1]\\ J\neq \emptyset}} (-1)^{\sum_{j\in J}y_j}\chi\left(\sum_{j\in J}\sum_{i=1}^e \gamma_i \alpha^{t_i j}x^{t_i}\right)\right) \\
      & =  \frac{1}{2^s} \left( 2^m-1+ \sum_{\substack{J\subseteq [s-1]\\ J\neq \emptyset}}(-1)^{\sum_{j\in J}y_j} \sum_{x\in \F_{2^m}^*} \chi\left(\sum_{i=1}^e \gamma_i \sum_{j\in J}\alpha^{t_i j}x^{t_i}\right)\right)
    \end{split}
    \end{equation*}

    Thus, we have
    \[
    \begin{split}
        \left|N-\frac{2^m-1}{2^s}\right|\leq \frac{1}{2^s} \sum_{\substack{J\subseteq [s-1]\\ J\neq \emptyset}} \left|\sum_{x\in \F_{2^m}^*} \chi\left(\sum_{i=1}^e \gamma_i \sum_{j\in J}\alpha^{t_i j}x^{t_i}\right)\right|
    \end{split}
    \]
    
    Unless $J=\emptyset$, $\sum_{j\in J}\alpha^{t_i j}\neq 0$ for every $i$, for otherwise, some $g_i$ would divide $J(X)\eqdef\sum_{j\in J}X^j$, which would imply $m\leq \deg J\leq s-1$, a contradiction. Since the sequence is not constant zero, some $\gamma_i$ is non-zero, and we conclude the polynomial $\sum_{i=1}^e \gamma_i \sum_{j\in J}\alpha^{t_i j}x^{t_i}$ is non-zero (here we are also using the fact that all $t_i$'s are different). Furthermore, the degree of the polynomial is equal to one of the numbers $t_i$, which are odd, and thus coprime with the size of the field $2^m$. The degree is at most $\max t_i$, so by the Weil-Carlitz-Uchiyama bound 
    \[ 
    \left| \sum_{x\in \F_{2^m}^*} \chi\left(\sum_{i=1}^e \gamma_i \sum_{j\in J}\alpha^{t_i j}x^{t_i} \right)\right| \leq (\max_i t_i-1)2^{m/2}+1,
    \]
     and the desired bound follows.
\end{IEEEproof}

\begin{remark}
In Theorem~\ref{thm:patternupperbound}, we note that if $\max_{i=1,\dots,e}t_i<3$, then $t_i=1$ for all $i$, and necessarily $e=1$ and the sequence is a PN sequence, in which case every non-zero pattern of length $m$ is guaranteed to occur exactly once.
\end{remark}

\begin{corollary}\label{cor:patternupperbound}
Consider the setting of Theorem~\ref{thm:patternupperbound}, and any pattern $y\in \F_2^s$ of length $s$. If $\max_{i=1,\dots,e}t_i>1$ and
\[
        s \leq\frac{m}{2}-\log_2\left(\max_{i=1,\dots,e} t_i-1\right),
\]
    then any non-zero LFSR sequence with connecting polynomial $g$ as in Theorem \ref{thm:patternupperbound} contains the pattern $y$.
\end{corollary}

\begin{IEEEproof}
    From~\eqref{eq:thmpattern}, we can lower bound $N$ by
    \[
    N\geq \frac{1}{2^s}(2^m-1-(2^s-1)(1+(\max t_i -1)2^{m/2})).
    \]
    This is greater than $0$ if and only if
    \[
    s<\log_2\parenv*{1+\frac{2^m-1}{1+(\max t_i -1)2^{m/2}}}.
    \]
    It is a routine computation to verify that if $\max t_i\geq 3$
    \[
    1+\frac{2^m-1}{1+(\max t_i -1)2^{m/2}}> \frac{2^{m/2}}{(\max t_i-1)},
    \]
    and so it suffices to ask
    \[
    s\leq \frac{m}{2}-\log_2 (\max t_i -1),
    \]
    for the number of occurrences of $y$ to be greater than $0$.
\end{IEEEproof}

By employing the generalized version of the Weil-Carlitz-Uchiyama bound found in Corollary \ref{coro:generalweilconsequence}, we can extend Theorem \ref{thm:patternupperbound} as follows:

\begin{theorem}\label{thm:melaspatternbound}
Consider a connection polynomial $g(X)=\prod_{i=1}^e g_i(X) \cdot \prod_{i=1}^d h_i(X)$, where $g_i\in \F_2[X]$, $1\leq i\leq e$, and $h_j\in \F_2[X]$, $1\leq j\leq d$, are distinct irreducible polynomials of the same degree $m$. Let $\alpha\in \F_{2^m}$ be a primitive element, and let $t_i,u_i\geq 1$ be odd positive integers such that $\alpha^{t_i}$ is a root of $g_i$ for every $1\leq i\leq e$, and $\alpha^{-u_i}$ is a root of $h_i$ for every $1\leq i\leq d$. Under these conditions, the following holds:

    Consider any pattern $y\in \F_2^s$ of length $s\leq m$. Then, the number of occurrences $N$ of $y$ in a window of length $2^m-1$ of any non-zero LFSR sequence with connection polynomial $g$ as above, satisfies
    \[
    \abs*{N-\frac{2^m-1}{2^s}}\leq \parenv*{1-\frac{1}{2^s}} (\max t_i +\max u_i)2^{m/2}.
    \]
\end{theorem}

\begin{IEEEproof}
The proof is the same as that of Theorem~\ref{thm:patternupperbound}, but Corollary~\ref{coro:generalweilconsequence} is used instead of the Weil-Carlitz-Uchiyama bound. In case the minimal polynomial of the sequence does not contain any factors $g_i$ (or $h_i$), the problem is reduced to the case of Theorem \ref{thm:patternupperbound}.
\end{IEEEproof}

\begin{remark}
In Theorem~\ref{thm:melaspatternbound}, notice that if $g$ is not the minimal polynomial, then the upper bound might be an overestimate. For example, if the initial condition coefficients $\gamma$ corresponding to the roots with negative exponents $-u_i$ are all zero, then the bound in Theorem \ref{thm:patternupperbound} applies.
\end{remark}

As a corollary, we obtain:
\begin{corollary}\label{coro:melaspattern}
Consider the setting of Theorem~\ref{thm:melaspatternbound}, and any pattern $y\in \F_2^s$ of length $s$. If
    \[
    s\leq \frac{m}{2}-\log_2(\max t_i+\max u_i),
    \]
    then any non-zero LFSR sequence with connecting polynomial $g$ as in Theorem \ref{thm:melaspatternbound} contains the pattern $y$.
\end{corollary}

\begin{IEEEproof}
    The proof is completely analogous to that of Corollary \ref{cor:patternupperbound}.
\end{IEEEproof}

\subsection{Improved bounds on the burst-covering radius}
We are now able to tighten the bounds in Theorem \ref{thm:basicboundcyclic}, and under certain conditions we can give the exact value of the burst-covering radius of cyclic codes: 

\begin{theorem}\label{thm:threeparts}
Consider an $[n,n-r]_2$ binary cyclic code, $\cC$, with generator polynomial $g= \prod_{i=1}^e g_i$, where $g_i\in\F_2[X]$, for $1\leq i\leq e$, are distinct irreducible factors of degrees $d_1\leq d_2\leq \dots \leq d_e$, respectively, and $r=\sum_{i=1}^e d_i$. Denote the burst-covering radius of $\cC$ by $b$. Then, the burst-covering radius $b$ of $\cC$ satisfies: 
    \begin{itemize}
        \item [1)] If $g_1$ is non-primitive,  $b\geq r-d_1+2$. 
        \item[2)]The burst-covering radius is upper-bounded by

        \[
        b\leq r- \min_{J\subseteq \set{1,\dots,e}, J \neq \emptyset} \varphi(J),
        \]
        where 
        \[
        \varphi(J)=\begin{cases}
            d_j-1 & \text{ if } J=\{j\} \text{ and } g_j \text{ is primitive}\\
            \floor{\log_2(\lcm(\ord(g_j):j\in J))-\sum_{j\in J}d_j/2} & \text{ otherwise}
        \end{cases}
        \]
        \item[3)] In the case $e=2$, suppose both $g_1$ and $g_2$ are primitive, $d_1<d_2$, and that either $\gcd(d_1,d_2)<d_2-d_1$, or $d_2-d_1\leq 2$. Then $b=r-d_1+1=d_2+1$. 
        
    \end{itemize}
\end{theorem}

\begin{IEEEproof}
1) Consider all possible LFSR sequences with connection polynomial $g_1$ (in particular, they also have connection polynomial $g$). If $g_1$ is not primitive, then there are at least two rotationally non-equivalent non-zero such sequences. The pattern $0\dots01$ (with $d_1-1$ zeros) can only appear in one of them (since the order of the LFSR is $d_1$), which means there exists a sequence with connection polynomial $g$ containing no runs of zeros of length $d_1-1$. By equation \eqref{eq:bfinalequivdefinition}, $b>r-d_1+1$.\\

2) An LFSR sequence with connection polynomial $g$ has minimal polynomial $\prod_{j\in J}g_j$ for some $J\subseteq \set{1,\dots, e}$. The minimal period is then equal to the least common multiple of the orders of $g_j$ for $j\in J$, $\pi = \lcm\set{\ord(g_j): j\in J}$. 
        
 If $J=\{j\}$ for some $j$ and $g_j$ is primitive, then the sequence with connection polynomial $g_j$ is a PN sequence, and its maximum run of zeros has length $d_j-1$. 
 
 Now suppose $J$ is not a singleton corresponding to a primitive polynomial. If $s\leq \min_{j\in J}d_j$, then by Theorem \ref{thm:niederreiterfreq} the number $N$ of runs of $s$ consecutive zeros in such a sequence satisfies
        \[
        2^sN\geq \lcm\set*{\ord(g_j): j\in J} -(2^s-1)2^{\sum_{j\in J}d_j/2}.
        \]
    This is greater than $0$ if and only if
    \[
    s<\log_2\left(1+\frac{\lcm\set{\ord(g_j): j\in J}}{2^{\sum_{j\in J}d_j/2}}\right)
    \]
    Thus, it suffices to ask that $s\leq \floor{\log_2(\lcm\set{\ord(g_j):j\in J})-\sum_{j\in J}d_j/2}$ to guarantee that the sequence contains a run of $s$ zeros. 
    
    Notice that $\varphi(\{1\})< d_1 =\min_{1\leq i\leq e} d_i$ and we do not need to include the condition $s\leq \min_{j\in J}d_j$ explicitly in the formula.\\

3) Thanks to~\eqref{eq:bfinalequivdefinition}, it suffices to show that any LFSR sequence with connection polynomial $g$ contains a run of at least $d_1-1$ zeros. Due to the primitivity assumption, an LFSR sequence with minimal connection polynomial $g_i$, for $i=1$ or $i=2$, is a PN sequence. Thus, it contains a run of $d_i-1\geq d_1-1$ zeros. 
    
    If the minimal connection polynomial is $g$, then by Theorem \ref{thm:niederreiterfreq} as in 2) it suffices to check
    \begin{equation}\label{eq:proofthmattainlowerbound}
    d_1-1< \log_2\left(1+\frac{\lcm(\ord(g_1),\ord(g_2))}{2^{(d_1+d_2)/2}}\right).    
    \end{equation}
    We know $\ord(g_1)=2^{d_1}-1$ and $\ord(g_2)=2^{d_2}-1$. Notice
    \[
\lcm(2^{d_1}-1,2^{d_2}-1)=\frac{(2^{d_1}-1)(2^{d_2}-1)}{2^{\gcd(d_1,d_2)}-1},
    \]
    where we have used that $\gcd(2^a-1,2^b-1)=2^{\gcd(a,b)}-1$ \footnote{That the RHS divides the LHS is straightforward. If we let $d=\gcd(a,b)$ and $d'=\gcd(2^a-1,2^b-1)$, then $d=ax+by$ for some integers $x$ and $y$. Since $2^a \equiv 1 \pmod{d'}$ and $2^b\equiv 1 \pmod{d'}$, we have $2^d = 2^{ax+by}\equiv 1\pmod {d'}$, so the LHS divides the RHS.}. By Lemma \ref{lemma:horrible} in the Appendix,
    \[
    1+\frac{(2^{d_1}-1)(2^{d_2}-1)}{(2^{\gcd(d_1,d_2)}-1)2^{(d_1+d_2)/2}}>2^{(d_1+d_2)/2-\gcd(d_1,d_2)},
    \] 
    so it suffices to ask 
    \[
    d_1-1\leq (d_1+d_2)/2-\gcd(d_1,d_2).
    \]
    If $\gcd(d_1,d_2)<d_2-d_1$, then $\gcd(d_1,d_2)\leq \frac{d_2-d_1}{2}$, and the previous inequality holds. If $\gcd(d_1,d_2)=d_2-d_1$, then the above inequality holds if and only if
    \[
    d_1-1\leq 3d_1/2-d_2/2 \Longleftrightarrow d_2-d_1\leq 2,
    \]
    and the claim is proved.
\end{IEEEproof}

Theorem \ref{thm:threeparts} 1) shows that the immediate lower bound on the covering radius from Theorem \ref{thm:basicboundcyclic} cannot be attained unless the lowest-degree irreducible factor in the generator polynomial for the code is primitive. On the other hand, 3) shows that there are codes which attain this lower bound. It is interesting to note that the conditions in 3) are sufficient, but need not be necessary. In fact, it might be possible that some stronger version of~\eqref{eq:niederreiterfreq} would allow to remove the conditions on $\gcd(d_1,d_2)$ altogether, as we do not have any example of primitive polynomials $g_1$ and $g_2$ for which $b>r-d_1+1$.

\section{Burst-Covering Radius of Long BCH Codes and Melas Codes}
\label{sec:bch}

Recall the definition of the binary primitive BCH code of length $n=2^m-1$ and designed distance $2e+1$: given $\alpha\in \F_{2^m}$ primitive, the parity-check matrix is defined as
\[
H = \begin{pmatrix}
    1 & \alpha & \alpha^2 & \dots & \alpha^{n-1}\\
    1 & \alpha^3 & \alpha^6 & \dots  & \alpha^{3(n-1)}\\
    \vdots & \vdots & \vdots & \ddots & \vdots \\
    1 & \alpha^{2e-1} & \alpha^{2(2e-1)} & \dots & \alpha^{(2e-1)(n-1)}
\end{pmatrix}.
\]
The parity-check matrix of a binary primitive Melas codes is given by
\[
H = \begin{pmatrix}
    1 & \alpha & \alpha^2 & \dots & \alpha^{n-1}\\
    1 & \alpha^{-1} & \alpha^{-2} & \dots  & \alpha^{-(n-1)}
\end{pmatrix}.
\]
We denote the two codes as $\BCH(e,m)$ and $\Melas(m)$. In what follows, we will consider BCH codes such that 
\begin{equation}\label{eq:longBCH}
    2^{\lceil m/2\rceil}> 2e-1,
\end{equation}
which guarantees that:

\begin{lemma}[{{\cite[Chapter 9 Theorem 7]{MacSlo78}}}]\label{lemma:basicpropslongBCH}
    Let $M_1, M_3, \dots, M_{2e-1}$ be the minimal polynomials of $\alpha, \alpha^3, \dots, \alpha^{2e-1}$ respectively. If  \eqref{eq:longBCH} is satisfied, $\deg(M_i)=m$ for all $i=1,3,\dots, 2e-1$, and $M_1, M_3, \dots, M_{2e-1}$ are pairwise coprime. 
\end{lemma}

Lemma \ref{lemma:basicpropslongBCH} implies that under condition~\eqref{eq:longBCH}, the generator polynomial of $\BCH(e,m)$ is 
\[
M(X)\eqdef \prod_{i=1}^e M_{2i-1}.
\]
We can now give an upper bound on the burst-covering radius of $\BCH(e,m)$:
\begin{theorem}\label{thm:upperbound}
    If $e\geq 2$, the burst-covering radius $b$ of $\BCH(e,m)$ which satisfies \eqref{eq:longBCH}, is bounded by
    \[
   b \leq \begin{cases}
        m\parenv*{e-\frac{1}{2}}+\ceil{\log_2(e-1)}+1 & \text{ if } 2\mid m\\
        m\parenv*{e-\frac{1}{2}}+\ceil{\log_2(e-1)+\frac{1}{2}}+\frac{1}{2} & \text{ if } 2\nmid m
    \end{cases}
     .
    \]
\end{theorem}

\begin{IEEEproof}
Instantiating Corollary \ref{cor:patternupperbound} with the generator polynomial of $\BCH(e,m)$, $M(X)=\prod_{i=1}^e M_{2i-1}$, we can take $t_i=2i-1$ for $1\leq i\leq e$. We deduce that for any initial condition, an LFSR sequence with connection polynomial $M$ will contain a run of zeros of any length up to $\lfloor m/2 -\log_2(2e-2)\rfloor$. Thus, $Z(M,(a_0,\dots,a_{me-1}))\geq \lfloor m/2 -\log_2(2e-2)\rfloor$ for any $(a_0,\dots,a_{me-1})$. From \eqref{eq:bfinalequivdefinition} we deduce that
    \[
    b \leq me- \lfloor m/2 -\log_2(2e-2)\rfloor.
    \]
    The claim follows by a straightforward analysis according to the parity of $m$.
\end{IEEEproof}

We can obtain a similar result for Melas codes:

\begin{theorem}
\label{thm:melasupper}
    The burst-covering radius $b$ of $\Melas(m)$, $m\geq 3$, satisfies
    \[
    b\leq \begin{cases}
        \frac{3}{2}m +1 & \text{ if } 2\mid m\\
        \frac{3}{2}m+\frac{3}{2} & \text{ if }2\nmid m
    \end{cases}.
    \]
\end{theorem}

\begin{IEEEproof}
    When $m\geq 3$, we have that $\alpha$ and $\alpha^{-1}$ are not conjugates, and therefore their respective minimal polynomials $M_1$ and $\overleftarrow{M_1}$ are distinct, where $\overleftarrow{h}=X^{\deg(h)}h(X^{-1})$. The proof follows from Corollary \ref{coro:melaspattern} by noting that the generator polynomial for the code is $g=M_1\cdot \overleftarrow{M_1}$. Thus, $e=d=1$ and $u_1=t_1=1$.
\end{IEEEproof}

\begin{table}
\caption{The exact burst-covering radius of $\BCH(2,m)$, $\Melas(m)$, and the upper bounds of Theorems~\ref{thm:upperbound} and \ref{thm:melasupper}}
\label{tab:exact}
\begin{center}
\begin{tabular}{c|cccc}
\hline\hline
    $m$ & BCH & Melas & Upper Bound & $M_1(X)$ \\
    \hline 
    6 & 9  & 10 & 10 & $X^6+X^4+X^3+X+1$\\
    7 & 11 & 11 & 12 & $X^7+X+1$\\
    8 & 12 & 12 & 13 & $X^8+X^4+X^3+X^2+1$\\
    9 & 13 & 14 & 15 & $X^9+X^4+1$\\
    10 & 14 & 15 & 16 & $X^{10}+X^6+X^5+X^3+X^2+X+1$\\
    11 & 16 & 16 & 18 & $X^{11}+X^2+1$ \\
    12 & 17 & 17 & 19 & $X^{12} + X^7 + X^6 + X^5 + X^3 + X + 1$\\
    13 & 18 & 18 & 21 & $X^{13} + X^4 + X^3 + X + 1$\\
    14 & 19 & 20 & 22 & $X^{14} + X^7 + X^5 + X^3 + 1$\\
    \hline\hline
\end{tabular}    
\end{center}
\end{table}

We can compute the actual value of the burst-covering radius for small codes by using a computer program. For example, for $\BCH(2,m)$ and $\Melas(m)$, the results are given in Table~\ref{tab:exact}. Notice that when $e=2$, the upper bounds for both codes coincide. The burst-covering radius can depend on the primitive root $\alpha$ chosen to define the code, and thus we include the chosen $M_1(X)$. 

The gap between the lower bound from Theorem \ref{thm:basicboundcyclic} and the previous upper bounds raises the question of whether the lower bound can be improved. Proving a lower bound would require a result on the existence of a (non-zero) LFSR sequence with a connecting polynomial as in Theorem \ref{thm:patternupperbound} that does \emph{not} contain a certain run of zeros (or more in general, does not contain a certain window pattern). 

We are not aware of any result of this type in the literature, and the techniques we have used so far do not seem to be effective for approaching this problem. However, we can give a slight improvement over the lower bound $b\geq (e-1)m+1$ from Theorem \ref{thm:basicboundcyclic}:

\begin{theorem}
\label{thm:bchmelaslower}
If $e\geq 2$, the burst-covering radius $b$ of $\BCH(e,m)$ which satisfies \eqref{eq:longBCH}, is bounded below by
    \[
    b\geq (e-1)m+2.
    \]
    Similarly, $b\geq m+2$ for $\Melas(m)$, $m\geq 3$.
\end{theorem}
\begin{IEEEproof}
    Since $e\geq 2$, $\BCH(e,m)$ is not the Hamming code and $b\geq 2$. According to inequality \eqref{eq:bound1}, $2^m-1\geq 2^{em-b+1}$ for the said BCH code. If $b=(e-1)m+1$, this implies $2^m-1\geq 2^m$, a contradiction. The same argument proves the bound for Melas codes.
\end{IEEEproof}

\section{Burst-Covering Algorithm for Binary Cyclic Codes}
\label{sec:algo}

Given a $b$-burst-covering code it is of interest to design an efficient algorithm which, given a syndrome $x\in \F_2^r$, returns a linear combination of $b$ consecutive columns equal to $x$. 

For binary cyclic codes, Algorithm \ref{alg:covering} is a natural consequence of the upper bound in Theorem \ref{thm:basicboundcyclic}, and Lemma \ref{lemma:fxkg}\footnote{Algorithm \ref{alg:covering} can be thought of as an adaptation of the ``burst trapping'' algorithm for decoding burst-error-correcting codes to the burst-covering setting, see \cite[Chapter 8]{McEl04}.}. Lines \ref{line:findf1} through \ref{line:findf3} find a pattern $f$ which generates $x$, using the fact that the first $r$ columns of $H$ are guaranteed to generate every possible syndrome. We have that $\LC(0,f)=x$. The While loop finds a suitable power $t$ such that $X^t f\pmod g$ has degree less than $b$ (or some desired threshold $b'$). The resulting polynomial $\hat{f}$ generates the same linear combinations as the original pattern $f$, but is shifted by $t$: $\LC(i,\hat{f})=\LC(i+t,f)$, so $\LC(-t \pmod n,\hat{f})=x$.

As an alternative for lines \ref{line:findf1}-\ref{line:findf3}, any covering algorithm (in the traditional sense) for the code can be used to find an initial pattern $f$, which should then be reduced modulo $g$.

\begin{algorithm}[t]
\caption{Burst-covering algorithm for cyclic codes}\label{alg:covering}
\begin{algorithmic}[1]
\State \textbf{Input:} $H\in \F_2^{r\times n}$, $g$ \text{ generator polynomial}, $x\in \F_2^{r}$, $b'\geq b$
\State \textbf{Output:} $f\in \cF_{b'}$, $0\leq i\leq n-1$ \text{ such that } $x=\LC(i,f)$
\Function{Burst\_cover}{$H, g, x, b'$}
\If{$x=0$}
    \State \Return $(0,0)$ 
\EndIf
\State $A\gets H[0 \dots r-1][0 \dots r-1]$ \label{line:findf1}
\State $y\gets A^{-1}x$ \label{line:findf2}
\State $f\gets \sum_{i=0}^{r-1}y_iX^i$ \label{line:findf3}
\State $t\gets 0$
\While{$\deg(f)\geq b'$}
    \State $f\gets (X\cdot f\pmod g)$
    \State $t\gets t+1$
\EndWhile
\State $i\gets -t \pmod n$
\State \Return $(i,f)$
\EndFunction
\end{algorithmic}
\end{algorithm}

The time complexity of Algorithm \ref{alg:covering} is $\cO(r^3+r\cdot n)$: the term $r^3$ corresponds to solving the linear system in line \ref{line:findf2}; the second is the maximum number of iterations of the while loop (which is $\leq\ord(g)\leq n$) times $r$, the degree of $g$, which determines the time required for the polynomial addition. In the case of $\BCH(e,m)$ codes, the time complexity becomes $\cO(e^3 \log^3 n+ e\cdot n\log n)$. Notice that $A^{-1}$ can be pre-computed when working with a fixed code, and thus the cost per syndrome is $\cO(r^2+r\cdot n)$.

Algorithm \ref{alg:computeradius} computes the burst-covering radius of a cyclic code with simple roots based on the characterization \eqref{eq:bfinalequivdefinition}. The function $\texttt{run\_LFSR}$ runs the LFSR with connection polynomial $g$ and initial state $\texttt{initial\_state}$ while keeping track of the longest run of zeros in the LFSR sequence and updating the $\texttt{used\_LFSR\_states}$ in place. If we denote by $\ell_v$ the length of the LFSR sequence with initial state $v$, then the time complexity of $\texttt{run\_LFSR(v,g,used\_LFSR\_states)}$ is $\cO(\ell_v\cdot r)$. Thus, the total time complexity is $\cO\parenv*{r\cdot\sum_{v}\ell_v}$, where the sum runs over the initial windows, one for each distinct (rotationally non-equivalent) LFSR sequence. Thus, the time complexity is $\cO(r2^r)$ while the space complexity is $\cO(2^r)$. In the case of $\BCH(e,m)$ codes this is $\cO(e n^e\log n)$.

\begin{algorithm}[t]
\caption{Computing the burst-covering radius of cyclic codes}\label{alg:computeradius}
\begin{algorithmic}[1]
\State \textbf{Input:} $g$ \text{ generator polynomial of degree } $r$
\State \textbf{Output:} $b$ 
\Function{Compute\_Burst\_radius}{$g$}
    \State $\texttt{min\_Z}\gets \infty$
    \State $\texttt{used\_LFSR\_states}\gets \underbrace{[\texttt{False},\dots, \texttt{False}]}_{2^r}$
    \For{\texttt{initial\_state = [0\dots 01],\dots,[1\dots 11]} }
        \If{$\texttt{not used\_LFSR\_states[initial\_state]}$}
            \State $\texttt{Z}\gets \texttt{run\_LFSR(initial\_state, g, used\_LFSR\_states)}$
            \State $\texttt{min\_Z}\gets \texttt{min(min\_Z, Z)}$
        \EndIf
    \EndFor
    \State \Return $\texttt{r-min\_Z}$
\EndFunction
\end{algorithmic}
\end{algorithm}

The previous algorithm can be considerably slower than Algorithm 2 in \cite{Matt1980DeterminingTB}, which determines the burst-error-correcting limit of a cyclic code in time $\cO(kr^2)$, where $k$ stands for the dimension of the code. It would be of interest to determine if the time can be improved in the covering case.

\section{Connection to the critical exponent}\label{sec:criticalexp}

Recall the definition of the critical exponent of a linear code:
\begin{definition}[\cite{greene1976weight}]
 The critical exponent of an $[n,n-r]_q$ code $\cC\subseteq \F_q^n$ is the minimum number $c$ such that $\cC$ contains $c$ codewords whose supports cover all $n$ coordinate positions. If $\cC$ has a coordinate which is always zero, the critical exponent is defined to be $\infty$.
\end{definition}

The critical exponent of a code is related to the famously hard critical problem for matroids, posed by Crapo and Rota \cite{crapo1970foundations}. For binary cyclic codes, we can give a relatively straightforward upper bound for the critical exponent based on the burst-covering radius of the dual code:

\begin{theorem}\label{thm:critexpbound}
    Given $\cC$ a binary $[n,n-r]$ cyclic code with burst-covering radius $b$, $\cC\neq \F_2^n$, the critical exponent of its dual code satisfies:
    \[
        c(\cC^{\perp})\leq r-b+1.
    \]
\end{theorem}

\begin{IEEEproof}
    By Corollary \ref{coro:equivdefconseczeros}, there is a codeword $c\in\cC^{\perp}$ such that its maximum run of zeros has length $r-b$. Then the union of the supports of $c$ and its $r-b$ shifts to the right $X\cdot c\pmod{X^n-1}, X^2\cdot c\pmod{X^n-1}, \dots, X^{r-b}\cdot c\pmod{X^n-1}$ is $\{1,\dots, n\}$.
\end{IEEEproof}

We can instantiate Theorem~\ref{thm:critexpbound} to the case of BCH and Melas codes to obtain the following corollary:

\begin{corollary}
\label{cor:critbchmelas}
For $m$ and $e$ satisfying~\eqref{eq:longBCH}, $e\geq 2$, 
\[
c(\BCH(e,m)^\perp) \leq m-1.
\]
For $m\geq 3$, 
\[
c(\Melas(m)^\perp) \leq m-1.
\]
\end{corollary}
\begin{IEEEproof}
Simply combine Theorem~\ref{thm:critexpbound} together with the lower bound from Theorem~\ref{thm:bchmelaslower}.
\end{IEEEproof}

These last two bounds can also be recovered by employing Bose-Burton's Theorem (see \cite{BoBu66}). However, Table \ref{tab:exact} suggests that these bounds are not tight. Notice that an improvement on the lower bound from Theorem \ref{thm:bchmelaslower} will immediately yield improvements on the previous upper bounds.

\section{Conclusion}
\label{sec:conc}

In this paper we initiated the study over burst covering codes. We focused in particular on the rich family of cyclic codes. We showed how the burst-covering radius of cyclic codes depends on the length of runs of zeros in related LFSR sequences. In some cases we can give the exact burst-covering radius. In the specific case of BCH codes, we developed a new bound on the existence of patterns in LFSR sequences, complementing those already found in the literature. For binary BCH codes and Melas codes it appears that the upper bound on the burst-covering radius is close to the actual radius. We then provided an efficient burst-covering algorithm for cyclic codes. Finally, we showed how the burst-covering radius of cyclic codes can help bound the critical exponent of these codes, whose computation is in general a notoriously difficult problem.

Many open questions remain. The immediate problem arising from the case BCH and Melas codes is the gap between the lower bound (Theorem~\ref{thm:bchmelaslower}) and the upper bound (Theorem~\ref{thm:upperbound} and Theorem~\ref{thm:melasupper}) on the burst-covering radius. The numerical results of Table~\ref{tab:exact} suggest the lower bound should be significantly improved. However, that requires results on the \emph{nonexistence} of patterns in LFSR sequences. To the best of our knowledge, these results are not known yet.

More generally, an important open question is to derive the burst-covering radius of other known families of codes, as well as developing efficient burst-covering algorithms for them. 

Finally, it is of interest to determine if bounds \eqref{eq:bound1} and \eqref{eq:bound3} can be improved or attained. We leave these questions for future work.

\textbf{Acknowledgments}
We acknowledge the support of the Natural Sciences and Engineering Research Council of Canada (NSERC), RGPIN-2026-06805 and ~CGRS D-611819-2026

Cette recherche a été financée par le Conseil de recherches en sciences naturelles et en génie du Canada (CRSNG), RGPIN-2026-06805 et ~CGRS D-611819-2026

\bibliographystyle{IEEEtranS}
\bibliography{allbib}

\appendix
\begin{lemma}\label{lemma:horrible}
    For all integers $a,b\geq 1$,
    \[
    1+\frac{(2^a-1)(2^b-1)}{(2^{\gcd(a,b)}-1)2^{(a+b)/2}}>2^{(a+b)/2-\gcd(a,b)}.
    \]
\end{lemma}

\begin{IEEEproof}
    Let $c=\gcd(a,b)$. The inequality holds if and only if
    \[
    \frac{(2^c-1)2^{(a+b)/2}+(2^a-1)(2^b-1)}{(2^c-1)2^{(a+b)/2}}>\frac{2^{(a+b)/2}}{2^c} \Longleftrightarrow
    \]
    \[
    2^c(2^c-1)2^{(a+b)/2}+2^c2^{a+b}-2^{c}2^a-2^c2^b+2^c>2^{a+b}(2^c-1) \Longleftrightarrow
    \]
    \begin{equation}\label{eq:horribleineq}
    2^c(2^c-1)2^{(a+b)/2}+2^{a+b}+2^c>2^{a+c}+2^{b+c}    
    \end{equation}
    
    Suppose without loss of generality that $a\leq b$.
    
    We analyze different scenarios:
    \begin{itemize}
        \item If $a=1$: then $c=\gcd(a,b)=1$, and the inequality becomes
        \[
        2\cdot 2^{(b+1)/2}+2^{b+1}+2>4+2^{b+1}\Longleftrightarrow
        \]
        \[
        2^{(b+3)/2}>2\Longleftrightarrow (b+3)/2>1\Longleftrightarrow b>-1,
        \]
        so the inequality \eqref{eq:horribleineq} holds in this case.
        \item If $a>1$ and $c=\gcd(a,b)\leq a/2$ then 
        \[
        2^{a+c}+2^{b+c}\leq 2^{a+a/2}+2^{b+a/2}\leq 2\cdot 2^{b+a/2}
        \]
        If $a\geq 2$, then $a/2+1\leq a$, so this means $2^{a+c}+2^{b+c}\leq 2^{b+a/2+1}\leq 2^{a+b}$, and so \eqref{eq:horribleineq} holds.
        \item If $a>1$ and $c= \gcd(a,b)=a$, \eqref{eq:horribleineq} becomes:
        \[
        2^a(2^a-1)2^{(a+b)/2}+2^{a+b}+2^a>2^{2a}+2^{a+b}\Longleftrightarrow
        \]
       \[
       2^{3a/2+b/2}(2^a-1)>2^a(2^a-1)\Longleftrightarrow
       \]
       \[
       3a/2+b/2>a,
       \]
       which is true for positive $a,b$. 
    \end{itemize}
\end{IEEEproof}

\end{document}